\documentclass[twocolumn,showpacs,preprintnumbers,amsmath,amssymb]{revtex4}

\usepackage{epsfig} 
\usepackage{dcolumn}
\usepackage{bm}

\begin{document}

\preprint{APS/123-QED}

\title{Analysis of two-pathway coherent control \\ for precision measurement of weak optical transitions}

\author{D. Antypas$^{1}$ and D. S. Elliott$^{1,2}$}
\affiliation{%
  $^1$Department of Physics, $^2$School of Electrical and Computer Engineering \\ Purdue University, West Lafayette, IN  47907
}

\date{\today}

\begin{abstract}
We present a new technique based on two-pathway optical coherent control for the sensitive detection and precise measurement of highly-forbidden transitions in atomic systems.  Specifically, we show that ($\omega$, 2$\omega$) coherent control can be applied to the measurement of the magnetic dipole and electroweak parity nonconserving amplitudes in atomic cesium, with the principal advantage of reduced systematic errors related to field reversals often encountered in previous measurements of these effects.  We present a complete analysis in one specific geometry, and discuss prospects for improved laboratory determinations of these weak transition amplitudes.
\end{abstract}

\maketitle

\section{Introduction}
Highly-forbidden optical transitions have been measured with great precision using the interference between the forbidden transition and a much stronger transition between the same two states.  Key examples of these techniques are found in measurements of magnetic dipole transitions~\cite{BouchiatP76,ChuCC77,BouchiatGP80,BouchiatP82,GilbertWW84,BouchiatGP84,JacquierBPB86,StalnakerBDFY02}, electric quadrupole transitions~\cite{HunterWW86,MajumderT99}, and parity nonconserving (PNC) electroweak interactions~\cite{BucksbaumCH81,BouchiatGH82,BouchiatGPH84,DrellC85,GilbertNWW85,GilbertW86,VetterMMLF95,WoodBCMRTW97,GuenaLB05a}. Still, pushing the precision of these measurements to greater limits, as well as extending them to a variety of other atomic systems, is an active pursuit, in that it will allow further tests of the Standard model and of nucleon-nucleon interactions within the nucleus~\cite{HaxtonW01,GingesF04}.  Currently a number of measurements of PNC interactions in various systems are underway, employing a variety of different techniques, but each based on an interference between interactions driven by a single laser frequency component~\cite{GomezASOD07,DeMilleCMRK08,TsigutkinDFSYB09}.  Systematic effects associated with imperfect reversal of the various dc electric fields, magnetic fields, and optical polarization are often the limiting factor of these measurements.  In this letter, we report a new technique based upon coherent control concepts for measuring these extremely weak interactions, and show that it eliminates the requirement for field reversals as the primary means of shifting the phase between terms, thereby reducing these associated errors.

For concreteness, we consider measurement of the PNC-induced electric dipole moment using the $6s \: ^2S_{1/2} \rightarrow n^{\prime}s \: ^2S_{1/2}$ transition in atomic atomic cesium, but we expect the principles we discuss to be generally applicable to other cases as well.  For $n^{\prime}$ = 7, this transition was used extensively by Wieman {\it et al.}~\cite{WoodBCMRTW97} and Bouchiat {\it et al.}~\cite{GuenaLB05a} for the most precise measurements of PNC amplitudes (${\mathcal{E}}_{PNC}$) to date.  As an $L = 0 \rightarrow L=0$ transition, it is electric dipole and electric quadrupole forbidden.  Stark-induced electric dipole (St), magnetic dipole (M1), and PNC-induced electric dipole interactions are each linear in the applied optical field $\varepsilon^{\omega_1}$, where the frequency $\omega_1$ is resonant with one of the hyperfine components the $6s \:  \rightarrow n^{\prime}s $ transition.

The Stark-induced amplitude for transitions from state $|6s \: ^2S_{1/2}, F, m\rangle$ to the state $|n^{\prime}s \: ^2S_{1/2}, F^{\prime}, m^{\prime}\rangle$, where $F$ and $m$ ($F^{\prime}$ and $m^{\prime}$) represent the total angular momentum and its projection in the $\hat{z}$-direction of the $|6s\rangle$ ($|n^{\prime}s\rangle$) state, is, in the notation of Gilbert and Wieman~\cite{GilbertW86}, of the form
\begin{eqnarray} \label{eq:AStFull}
 A_{St}(F,m;F^{\prime},m^{\prime}) &=&  \\
      & &  \hspace {-1.2in} e^{i \phi^{\omega_1}} \left\{ \left[ \mbox{\rule[-0.3cm]{0cm}{0.6cm}} \alpha {\bf E} \cdot
        {\bf \varepsilon}^{\omega_1} \delta_{F,F^{\prime}} +  i \beta ( {\bf E} \times
        {\bf \varepsilon}^{\omega_1} )_z C_{F,m}^{F^{\prime},m^{\prime}} \right]  \delta_{m,m^{\prime}} \right. \nonumber  \\
      & & \hspace {-1.2in} \left. + \beta \left[ \mbox{\rule[-0.3cm]{0cm}{0.6cm}} \pm i ({\bf E} \times {\bf \varepsilon}^{\omega_1})_x - ({\bf E} \times {\bf \varepsilon}^{\omega_1} )_y \right]  C_{F,m}^{F^{\prime},m^{\prime}} \delta_{m,m^{\prime} \pm 1} \right\} , \nonumber
\end{eqnarray}
where $\alpha$ and $\beta$ are the scalar and vector polarizabilities, ${\bf \varepsilon}^{\omega_1}$ and $\phi^{\omega_1}$ are the electric field amplitude and phase of the optical field, and the $C_{F,m}^{F^{\prime},m^{\prime}}$ are constants derived from the Clebsch-Gordon coefficients~\cite{GilbertW86}.  The polarizabilities are given in Ref.~\cite{BouchiatB75}.  The phase term in Eq.~(\ref{eq:AStFull}), $e^{i \phi^{\omega_1}}$, which can be omitted for single-beam excitation, must be retained in our analysis.

To first order, the $6s \: ^2S_{1/2} \rightarrow n^{\prime}s \: ^2S_{1/2}$ transition is magnetic dipole forbidden, but spin-orbit interactions and relativistic effects relax this somewhat, and magnetic dipole moments $M=\langle n^{\prime}S | \:  \mu_z/c \: | 6S \rangle$, where $\mu_z$ is the $z$-component of the magnetic dipole operator, have been determined for $n^{\prime}$ = 7~\cite{GilbertWW84,BouchiatGP84} and $n^{\prime}$ = 8~\cite{JacquierBPB86}.  The amplitude of the magnetic dipole transition is given by
 \begin{eqnarray} \label{eq:AM1Full}
  A_{M1}(F,m;F^{\prime},m^{\prime}) &=& e^{i \phi^{\omega_1}} \left\{ ({\bf {\hat{k}}} \times {\bf \varepsilon}^{\omega_1})_z  \delta_{m,m^{\prime}}  \right.  \\
   & &  \left. \hspace {-1.3in} +\left[\pm ({\bf {\hat{k}}} \times {\bf \varepsilon}^{\omega_1})_x +i ({\bf {\hat{k}}} \times {\bf \varepsilon}^{\omega_1})_y \right] \delta_{m,m^{\prime} \pm 1}
   \right\}  M C_{F,m}^{F^{\prime},m^{\prime}}, \nonumber
\end{eqnarray}
where ${\bf {\hat{k}}} = {\bf k}^{\omega_1}/|{\bf k}^{\omega_1}|$, and $ {\bf k}^{\omega_1}$ is the propagation vector of the $\omega_1$ beam.  Additionally, the electroweak interaction between the nucleus and the electrons, as well as higher order PNC effects, lead to a very slight mixing between even- and odd-parity eigenstates of the atom,  making an electric dipole transition slightly allowed as well.  The transition amplitude for this interaction is of the form
\begin{eqnarray}\label{eq:APNCFull}
  A_{PNC}(F,m;F^{\prime},m^{\prime}) &=& e^{i \phi^{\omega_1}} \left\{ \varepsilon_z^{\omega_1} \delta_{m,m^{\prime}} \right. \\
  & & \hspace {-1.2in} \left. + \left[ \pm \varepsilon_x^{\omega_1} + i \varepsilon_y^{\omega_1} \right] \delta_{m,m^{\prime} \pm 1} \right\}  i Im ({\mathcal{E}}_{PNC}) C_{F,m}^{F^{\prime},m^{\prime}}, \nonumber
\end{eqnarray}
where ${\mathcal{E}}_{PNC}$ is the matrix element for electric dipole transitions due to the state mixing by the PNC interactions.  To measure $A_{M1}$ or $A_{PNC}$ directly is problematic in that their magnitudes are typically well below the level of measurement noise.  Techniques using an interference between the weak transition and a stronger transition (the Stark-induced transition, for example) have therefore been developed to effectively amplify the signal to a detectable level.  For example, under conditions that allow a strong Stark-induced amplitude and a weak PNC amplitude on the same transition that add constructively, the net rate scales as $W_+ =|A_{St} + A_{PNC}|^2 \simeq |A_{St}|^2 + 2|A_{St}||A_{PNC}|$.  The interference between these various amplitudes can be reversed by reversing one or more of the fields that influences the sign of the amplitudes, resulting in a rate $W_- =|A_{St} - A_{PNC}|^2 \simeq |A_{St}|^2 - 2|A_{St}||A_{PNC}|$.  A precise measurement of the small difference between $W_+$ and $W_-$ can then be used to determine $|A_{PNC}|$.

In each of the previous measurements, a single laser field has been employed, and both interactions (strong and weak) are linear in the amplitude of this field.  We now consider this system under the influence of a second optical field composed of components at frequencies $\omega_2$ and $\omega_3$, where $\omega_2 + \omega_3 = \omega_1$, which is capable of driving the $6s \:  \rightarrow n^{\prime}s $ transition via a two-photon interaction.  In order for these amplitudes to interfere, the $\omega_1$ field component must be phase coherent with $\omega_2$ and $\omega_3$ components, as it will be when the former is generated from the latter using a nonlinear optical crystal for sum frequency generation.  We have previously demonstrated this interference between two-photon absorption and Stark-induced linear absorption on the cesium $6s \rightarrow 8s$ transition~\cite{GunawardenaE07}.  We write the transition amplitude for this interaction in a form similar to that of the Stark-induced transition given by Eq.~(\ref{eq:AStFull}),
\begin{eqnarray} \label{eq:2PAFull}
 A_{2PA}(F,m;F^{\prime},m^{\prime}) &=&  e^{i (\phi^{\omega_2} + \phi^{\omega_3})} \times \\
      & &  \hspace {-1.3in}  \left\{ \left[ \mbox{\rule[-0.3cm]{0cm}{0.6cm}} \tilde{\alpha} {\bf \varepsilon}^{\omega_2} \cdot
        {\bf \varepsilon}^{\omega_3} \delta_{F,F^{\prime}} +  i \tilde{\beta} ( {\bf \varepsilon}^{\omega_2} \times
        {\bf \varepsilon}^{\omega_3} )_z C_{F,m}^{F^{\prime},m^{\prime}} \right] \delta_{m,m^{\prime}} \right. \nonumber \\
      & & \hspace {-1.3in}  + \left. \tilde{\beta} \left[\mbox{\rule[-0.3cm]{0cm}{0.6cm}} \pm i  ({\bf \varepsilon}^{\omega_2} \times {\bf \varepsilon}^{\omega_3})_x - ({\bf \varepsilon}^{\omega_2} \times {\bf \varepsilon}^{\omega_3} )_y \right]  C_{F,m}^{F^{\prime},m^{\prime}} \delta_{m,m^{\prime} \pm 1} \right\}, \nonumber
\end{eqnarray}
where ${\bf \varepsilon}^{\omega_2}$ and ${\bf \varepsilon}^{\omega_3}$ are the amplitudes, and phases $\phi^{\omega_2}$ and $\phi^{\omega_3}$, the phases, of the optical waves at frequencies $\omega_2$ and $\omega_3$, respectively, and the coefficients of the two-photon moments, $\tilde{\alpha}$ and $\tilde{\beta}$, are defined in a form similar to the Stark polarizabilities.  Bouchiat and Bouchiat~\cite{BouchiatB75} noted the relationship between Stark-induced transitions and two-photon absorption.  The interference between the two-photon amplitude and the amplitudes that are linear in $\varepsilon^{\omega_1}$ can be observed on $\Delta F = 0$ as well as $\Delta F = \pm1$ transitions.  The $\Delta F = 0$ transitions, however, present the following two advantages over $\Delta F = \pm1$ transitions:  (1) Systematic errors due to magnetic dipole contributions to $\mathcal{E}_{PNC}$ are smaller, and (2) the two frequencies $\omega_3$ and $\omega_2$ can be equal, so this measurement requires only a single laser source and the $\omega_1$ beam can be generated by frequency doubling the $\omega_2$ laser output in a nonlinear crystal. We will consider only $\Delta F = 0$, $\Delta m = 0$ transitions in the following, and write the two-photon transition amplitude of Eq.~(\ref{eq:2PAFull}) as $A_{2PA} = \tilde{\alpha} \left( \varepsilon^{\omega_2} \right)^2 e^{2i\phi^{\omega_2} }$.

To maintain a constant phase difference between the two-photon amplitude and the linear amplitudes, the optical beams must propagate in directions nearly co-linear with one another.  Without lose of generality, we define the $y$-axis along ${\bf {\hat{k}}}$, such that $\varepsilon_y^{\omega_1}$ must vanish for a plane wave or weakly-focussed beam. We allow an arbitrary static electric field ${\bf E} = E_x \hat{x} + E_y \hat{y} + E_z \hat{z}$, and consider a dc magnetic field ${\bf B}$ that is primarily in the $\hat{z}$-direction.  ${\bf B}$ will separate the various $m$-components: $\Delta E_{F,m} = \mu_B g_F mB_z$, where $\mu_B$ is the Bohr magneton and $g_F$ is 1/4 for F=4 and -1/4 for F=3.  The effect of the $\hat{x}$- and $\hat{y}$-components for $\bf B$, which can be present in an experiment due to imperfect alignment, is to mix the magnetic sublevels
\begin{eqnarray}\label{eq:ZeemanMixing}
    \overline{|ns \: ^2S_{1/2}, F, m\rangle} &=& |ns \: ^2S_{1/2}, F, m\rangle \\
    & & \hspace{-.5in} + |ns \: ^2S_{1/2}, F, m-1\rangle \frac{B_x + iB_y}{B_z}C_{F,m}^{F,m-1} \nonumber \\
    & & \hspace{-.5in} - |ns \: ^2S_{1/2}, F, m+1\rangle \frac{B_x - iB_y}{B_z}C_{F,m}^{F,m+1}. \nonumber
\end{eqnarray}
We include in this expression mixing among magnetic components of the same F, but omit mixing with other F states, an approximation that will be valid for the modest magnetic field strengths characteristic of these measurements.

\begin{figure}
  \includegraphics[width=8cm]{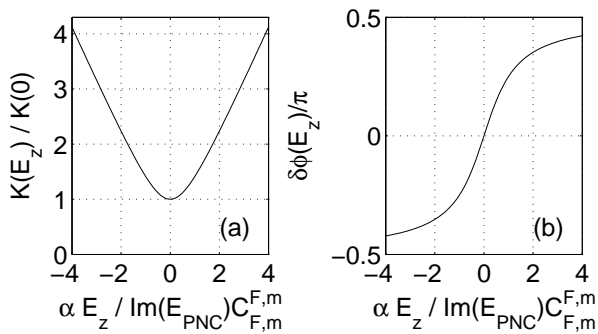}\\
  \caption{The (a) amplitude and (b) phase of the modulation signal vs $E_z$.}\label{fig:ampdphivsEz}
\end{figure}

We sum the four transition amplitudes,
\begin{eqnarray}\label{eq:Systematics}
\sum A & =  & A_{2PA}  +  \left[ \left\{ \mbox{\rule[-0.3cm]{0cm}{0.6cm}} \alpha  \: E_z \:  \varepsilon_z^{\omega_1} +  \alpha E_x \: \varepsilon_x^{\omega_1} - M \: \varepsilon_x^{\omega_1} C_{F,m}^{F,m} \right. \right. \nonumber \\
& & \hspace{-0.5 in} \left.  - M \: \varepsilon_z^{\omega_1} \frac{B_x}{B_z} \: \Delta C^{(2)} \right\} \: +  i \left\{ \mbox{\rule[-0.3cm]{0cm}{0.6cm}} Im ({\mathcal{E}}_{PNC}) \:  \varepsilon_z^{\omega_1} \: C_{F,m}^{F,m} \right. \nonumber \\
&   & \hspace{-0.5 in}  \:   - \beta E_y  \: \varepsilon_x^{\omega_1} \: C_{F,m}^{F,m}  + \beta E_y \: \varepsilon_z^{\omega_1} \: \frac{B_x}{B_z} \: \Delta C^{(2)} \\
& &  \hspace{-0.5 in}  + \beta E_z \: \varepsilon_x^{\omega_1} \: \frac{B_y}{B_z}\: \Delta C^{(2)}  \:\left.  \left. \hspace{0in}  - \beta E_x \:  \varepsilon_z^{\omega_1}  \: \frac{B_y}{B_z} \: \Delta C^{(2)} \:
 \right\} \: \right]  e^{i \phi^{\omega_1}},  \nonumber
\end{eqnarray}
where all terms except $\varepsilon_x^{\omega_1}$ are real, and
$\Delta C^{(2)} = \sum_{+/-} \left\{(C_{F,m}^{F,m\pm1})^2 - C_{F,m\pm1}^{F,m}C_{F,m}^{F,m\pm1} \right\}$ is 3/16 for F = 3, m = $\pm$3, and 1/4 for F = 4, m = $\pm$4.  This factor is small, but not negligible, in comparison to $C_{F,m}^{F,m} = \mp3/4$ for F = 3, m = $\pm$3 or $\pm1$ for F = 4, m = $\pm$4.
The relative scale of the different amplitudes in Eq.~(\ref{eq:Systematics}) depends on many factors, but for $E_y < $ 100 V/cm, the two-photon rate dominates all others, even with cw beam powers and modest focussing.  By carefully selecting beam polarizations and controlling static electric and magnetic fields, one can select a very limited number of contributions to the interaction.  For example, to determine the ratio $Im ({\mathcal{E}}_{PNC})/\alpha$,
 one can align the polarization of the harmonic beam ($\omega_1$) to the $\hat{z}$ direction, apply a low-magnitude dc field in the $\hat{z}$ direction, and zero the field components $B_x$, $B_y$, $E_x$, $E_y$, and $\varepsilon_x$.  In this geometry, the surviving transition moments sum to
\begin{equation}\label{eq:AII}
\sum A =  A_{2PA} +  \left( \alpha  E_z + i Im({\mathcal{E}}_{PNC}) C_{F,m}^{F,m} \right) \varepsilon_z^{\omega_1} e^{i \phi^{\omega_1}}.
\end{equation}
The transition rate scales as $W = |\sum A|^2$, yielding a transition rate of the form
\begin{equation}\label{eq:WII}
W =|A_{2PA}|^2 + K(E_z) \sin(\Delta \phi + \delta \phi(E_z)),
\end{equation}
where we have omitted the terms that are second-order in $\varepsilon_x^{\omega_1}$, and define $\Delta \phi = 2\phi^{\omega_2} - \phi^{\omega_1}$, the controllable phase difference between the optical fields.
The total transition rate of Eq.~(\ref{eq:WII}) has a large dc component that represents the two-photon absorption rate alone, and a weak sinusoidal modulation of amplitude   \begin{equation}\label{eq:Kpnc}
    K(E_z) = 2 |A_{2PA}| \sqrt{(\alpha E_z)^2 + (Im({\mathcal{E}}_{PNC} ) \; C_{F,m}^{F,m}  )^2} \; \varepsilon_z^{\omega_1}.
\end{equation}
This modulation is a result of the interference between the two-photon absorption from the $\omega_2$ beam and the sum of the various weak amplitudes (each of which are linear in the amplitude of the $\omega_1$ beam). We show the normalized amplitude $K(E_z)/K(0)$ and the phase shift $\delta \phi = \tan^{-1}(\alpha E_z/Im({\mathcal{E}}_{PNC}))$ as a function of $E_z$ in Fig.~\ref{fig:ampdphivsEz}.
Measurements of the modulation amplitude as a function of field $E_z$ can therefore be used to determine the ratio $|Im({\mathcal{E}}_{PNC})/\alpha|$.  At $E_z = 0$, the observed modulation is due solely to the interference between two-photon absorption and the PNC-induced amplitude, since the Stark-induced transition is absent in this case.  As $E_z$ is increased, the amplitude of the modulation increases due to the growing contribution of the Stark-induced transition.  The Stark- and the PNC-induced moments add in quadrature in Eq.~(\ref{eq:Kpnc}) since these amplitudes differ in phase by $\pi/2$, as shown in Eq.~(\ref{eq:AII}).
The scale of the variation in $E_z$ necessary to carry out these measurements is determined by $|Im({\mathcal{E}}_{PNC})/\alpha|$.  Using Wieman's results for $\mathcal{E}_{PNC} / \beta$~\cite{WoodBCMRTW97} on the $6s \rightarrow 7s$ transition, and for $\alpha/\beta$~\cite{ChoWBRW97}, we estimate $|Im({\mathcal{E}}_{PNC})/\alpha|$ to be $\sim$ 0.158 mV/cm.  A precise measurement as described above will therefore require controlled application of $E_z$ at this level, and cancelation of stray fields to a level significantly lower.  This can be achieved using precision voltage sources (which can control biases of field plates to the sub-$\mu$V level).  It is not necessary to calibrate the other factors that appear in Eq.~(\ref{eq:Kpnc}), {\it i.e.} $A_{2PA}$ or $\varepsilon_z^{\omega_1}$, in that one only needs to determine the relative variation in $K(E_z)$ vs. $E_z$.  The sign of $Im({\mathcal{E}})_{PNC}/\alpha$ can be determined from observations of $\delta \phi$ vs. $E_z$.

Stray fields $B_x$, $B_y$, $E_x$, $E_y$, and $\varepsilon_x^{\omega_1}$ can affect the amplitude of the modulation signal, and these must be minimized in order to make reliable measurements. We examine the individual terms of Eq.~(\ref{eq:Systematics}) in order to estimate their magnitude and suggest means for their reduction.  We first consider those terms that contribute to the imaginary part of this signal, i.e. those that will add to the $Im ({\mathcal{E}}_{PNC}) \:  \varepsilon_z^{\omega_1} \: C_{F,m}^{F,m}$ term directly.  We expect that the dominant term of this type that must be addressed is a magnetic dipole contribution, $- M \: \varepsilon_x^{\omega_1 \: \prime\prime} C_{F,m}^{F,m}$, where $\varepsilon^{\omega_1 \: \prime\prime}$ is the amplitude of the imaginary component of the harmonic field, $\varepsilon^{\omega_1} = \varepsilon^{\omega_1 \: \prime} + i \varepsilon^{\omega_1 \: \prime\prime}$.  We note that the Boulder experiments were also affected by a magnetic dipole contribution, which they countered by employing counter-propagating laser beams~\cite{WoodBCMRTW97}.  In their experiment, the $\varepsilon_x^{\omega_1 \: \prime\prime}$ field component was essential to their measurement, while in the scheme addressed here, it is not.  To reduce this component to the level of the PNC contribution requires $\varepsilon_x^{\omega_1 \: \prime\prime} \sim 0.5 \times 10^{-4} \: \varepsilon_z^{\omega_1} $.  Reduction to lower levels than this is, of course, highly desirable.  While this will clearly be an experimental challenge, recent progress using high-Q traveling-wave optical cavities to produce linearly-polarized light of high purity has been reported~\cite{SarafBK07}.  The degree to which this magnetic dipole term is reduced can be determined experimentally by reversing the direction of ${\bf B}$ and repeating the measurement of $K(E_z)$.  The spatially-varying ac Stark shifts present in the Boulder measurements will not be a factor in the present scheme.

Stray fields can also contribute to the apparent PNC signal through Stark-induced terms such as
$\beta E_y  \: \varepsilon_x^{\omega_1} \: C_{F,m}^{F,m}$. We estimate that this effect can be reduced to the 10$^{-3}$ level by reducing $E_y$ to $\sim$ 1 mV/cm and the polarization purity to $\varepsilon_x^{\omega_1 \: \prime} / \varepsilon_z^{\omega_1} < 10^{-3}$.  There are, of course, no detectors available to measure field strengths at these low levels; for this the methods described by Wood~\cite{Wood96} for nulling stray fields in their measurements should be applicable to the present measurements as well.  For example, referring to Eq.~(\ref{eq:Systematics}), increasing $B_x$ to elevate the $i \beta E_y \varepsilon_z^{\omega_1} (B_x/B_z) e^{i \phi^{\omega_1}} \Delta C^{(2)}$ term, allows one to null $E_y$, at which point signal modulation as a function of $\Delta \phi$ is minimized.  Other stray field components can be nulled in a similar manner.

To this point, we have discussed stray field contributions that are in-phase with the PNC term in Eq.~(\ref{eq:Systematics}).  There are also terms that contribute signal in phase with the $\alpha  \: E_z \:  \varepsilon_z^{\omega_1}$ term.  For two reasons, these appear to be of secondary importance, as long as they are constant in time and spatially homogeneous.  First, one can always apply a dc field $E_z$ large enough that $\alpha  \: E_z \:  \varepsilon_z^{\omega_1}$ is larger than all others by many orders.  Secondly, if some of these terms survive, their effect is to shift the apparent zero point of the applied field $E_z$.  Thus these in-phase contributions can be made negligible by measuring $K(E_z)$ over a sufficiently broad range of applied fields.

We conclude that two-pathway coherent control techniques can present a new means for determination of weak interaction amplitudes.  While we focussed our attention on measurement of $Im ({\mathcal{E}}_{PNC})$, the coherent control technique can also be used to measure the amplitude of the M1 transition moment using a different field geometry.  The primary advantage of the present technique is that the phase of the cross term interference signal can be varied by varying the optical phase difference between the laser beams.  While reversing the direction of dc electric fields, dc magnetic fields, or field polarizations can be used as a means of measuring and reducing stray field effects, phase variation for the primary measurements can be implemented through continuous, external, optical means such as a simple delay cell filled with a non-absorbing variable-pressure gas (argon or nitrogen, for example).  A delay cell is attractive in its simplicity and relative immunity from variations in polarization, amplitude, or beam alignment.  The measurement determines the magnitude and sign of $Im({\mathcal{E}}_{PNC})$ relative to $\alpha$, whose calculated value is more reliable than that of $\beta$.  Precise calibration of the two-photon amplitude is {\it not} required, since we are only interested in the relative variation in the modulation amplitude vs. $E_z$.  Our proposed measurements employ exclusively linear polarizations, a further simplification.  Finally, this technique can allow reduction of magnetic dipole contributions without introducing standing-wave optical fields.  Experimental efforts leading toward these measurements are currently underway in our laboratory.

Useful conversations with C. E. Wieman, M. A. Safronova, A. V. Smith, Yong P. Chen, and A. M. Weiner are gratefully acknowledged.

\end{document}